\documentclass[preprint,showpacs,preprintnumbers,amsmath,amssymb,nofootinbib]{revtex4}
\usepackage{booktabs}
\usepackage{mathrsfs}
\usepackage{epsfig}
\usepackage{graphicx}
\usepackage{dcolumn}
\usepackage{bm}
\usepackage{amsmath}
\usepackage{slashed}
\usepackage{multirow}

\let\jnfont=\rm
\def\NPB#1,{{\jnfont Nucl.\ Phys.\ B }{\bf #1},}
\def\PLB#1,{{\jnfont Phys.\ Lett.\ B }{\bf #1},}
\def\EPJC#1,{{\jnfont Eur.\ Phys.\ Jour.\ C }{\bf #1},}
\def\PRD#1,{{\jnfont Phys.\ Rev.\ D }{\bf #1},}
\def\PRL#1,{{\jnfont Phys.\ Rev.\ Lett.\ }{\bf #1},}
\def\MPLA#1,{{\jnfont Mod.\ Phys.\ Lett.\ A }{\bf #1},}
\def\JPG#1,{{\jnfont J.\ Phys.\ G}{\bf #1},}
\def\CTP#1,{{\jnfont Commun.\ Theor.\ Phys.\ }{\bf #1},}
\def\ZPC#1,{{\jnfont Z.\ Phys.\ C }{\bf #1},}
\def\JHEP#1,{{\jnfont JHEP \ }{\bf #1},}
\def\Rv{\not{\hbox{\kern-1pt $R$}}}
\def\p{\not{\hbox{\kern-3pt $p$}}}

\begin{document}

\title{Single top partner production in the Higgs to diphoton channel in the Littlest Higgs Model with $T$-parity}
\author{ Ning Liu$^{1}$}
\author{ Lei Wu$^{2}$}
\author{ Bingfang Yang$^{1}$}
\author{ Mengchao Zhang$^{3}$}
\affiliation{$^1$ College of Physics $\&$ Electronic Engineering, Henan Normal University, Xinxiang 453007, China\\
$^2$ ARC Centre of Excellence for Particle Physics at the Terascale, School of Physics, The University of Sydney, NSW 2006, Australia\\
$^3$ State Key Laboratory of Theoretical Physics, Institute of Theoretical Physics, Academia Sinica, Beijing 100190, China
   \vspace*{1.5cm} }%

\date{\today}

\begin{abstract}

The top partner as a hallmark of the Littlest Higgs model with $T$-parity (LHT model) has been extensively searched for during the Large Hadron Collider (LHC) Run-1. With the increasing mass limits on the top partner, the single production of the top partner will be dominant over the pair production. Under the constraints from the Higgs data, the electroweak precision observables and $R_b$, we find that the mass of $T$-even top partner ($T_+$) has to be heavier than 730 GeV. Then, we investigate the observability of the single $T$-even top partner production through the process $pp \to T_+ j$ with the sequent decay $T_+ \to th$ in the di-photon channel in the LHT model at the LHC. We find that the mass of $T_+$ can be excluded up to 800 GeV at $2\sigma$ level at 14 TeV LHC with the integrated luminosity ${\cal L}=3$ ab$^{-1}$.

\end{abstract}
 \maketitle

\section{INTRODUCTION}

The discovery of a 125 GeV Higgs boson at the LHC Run-1 \cite{higgs-atlas,higgs-cms} is a major step towards elucidating the electroweak symmetry breaking mechanism and marks a great triumph of the Standard Model (SM). However, without protection by a symmetry in the SM, the Higgs mass is quadratically sensitive to the cutoff scale $\Lambda$ via radiative corrections, rendering the theory with $m_h \ll \Lambda$ rather unnatural. This electroweak naturalness problem is widely considered as a major motivation for new physics beyond the SM.

Among many extensions of the SM, the Littlest Higgs with $T$-parity model (LHT model) is one of the most promising candidates that can successfully solve the electroweak naturalness problem \cite{lh,lht,lht-ewpos}. It is based on a nonlinear sigma model with a global $SU(5)$ symmetry, which is broken down to $SO(5)$ by a vacuum condensate $f \sim \Lambda/4\pi$ TeV. At the same time, the gauged subgroup $[SU(2)\times U(1)]^2$ is broken to its diagonal subgroup $SU(2)\times U(1)$ that is identified as the SM electroweak gauge group. All quadratically divergent one-loop contributions to the Higgs mass only first appear at two-loop level, but their values are suppressed by an additional loop factor and thus are sufficiently small to prevent the little hierarchy problem from being re-introduced at the TeV scale. Due to the implementation of the $T$ symmetry, the top quark sector is enlarged, which leads to the abundant phenomenology of top quark sector in the LHT model. In particular, the vector-like top partner that is related with the mechanism of canceling the large quadratic divergences in the Higgs mass from the SM top quark loop has been widely studied \cite{lht-tp}.

Since the top partner plays an important in understanding the electroweak naturalness problem, the ATLAS and CMS collaborations have performed the searches for the vector-like top partner through the pair or single production with different final states $bW$, $tZ$ and $th$ during the LHC Run-1. They have excluded the top partner with the mass less than about 700 GeV \cite{tp-atlas,tp-cms}. However, their bounds strongly depend on the assumptions on the decay branching ratios and the properties of the top partner, in particular its group representations. On the other hand, the indirect searches for the top partners through their contributions to the electroweak precision observables (EWPOs) \cite{lht-ewpos}, $Z$-pole observables \cite{lht-rb-1,lht-rb-2} and the flavor physics \cite{lh-flavor} have been extensively investigated. The non-observation of the top partners, in conjunction with the EWPOs and the recent discovery of a 125 GeV Standard Model-like (SM-like) Higgs boson, have tightly constrained the parameter space of the LHT model where the top partner can be light \cite{lht-tonini,lht-berger,lht-yang,lht-wang}. Theoretically, the LHT model with the top partner at TeV scale usually suffers from a higher degree of fine tuning. But phenomenologically, such a TeV top partner can be well probed at the LHC and future high energy colliders.

In this work, we will study the single $T$-even top partner ($T_+$) production through the process $pp \to T_+j$ with the sequent decay $T_+ \to th$ in the di-photon channel in the LHT model at the LHC. As the increasing mass limits on the top partner, the single top partner production will have the larger cross section than the pair production at the LHC, due to the collinear enhancement for the light quark emitting a $W$-boson in the high energy region \cite{willen}. In addition, the single fermionic top partner production has the unique event topology that offers a great opportunity to get rid of the large SM backgrounds. So the single top partner production is becoming more and more important at the LHC \footnote{In recent works \cite{tilman,wu}, the single stop production is also found to be a useful probe of the naturalness of the MSSM at the high luminosity LHC.}. The analyses of the singly produced top partners that decay to $bW$ and $tZ$ have been performed in Ref.~\cite{tp-single-1} and Ref.~\cite{tp-single-2}, respectively. Using the boosted object tagging methods, the authors in Ref.~\cite{tp-single-3} studied the search strategies of the single top partner production with the sequent decay $T^{'} \to th$ in various hadronic decay channels. In comparison with these exisiting studies, the $h \to \gamma\gamma$ channel usually has the small cross section but the very distinctive final states. Therefore, such a channel may become a complementary to the hadronic channels in the searches for the top partner.

This paper is organized as follows. In section II, we give a brief description of the top partner sector of the LHT model and perform a scan of the parameter space of the LHT model. Then we calculate the cross sections of the $T$-even top partner productions and its decay branching ratios in the allowed parameter space. In section III, we study the sensitivity of $T_+ (\to th) j$ production in diphoton channel at 14 TeV LHC. Finally, we draw our conclusions in section IV.

\section{Top partner in Littlest Higgs Model with T-parity}\label{section2}
The LHT model is a non-linear $\sigma$ model based on the coset space $SU(5)/SO(5)$. For the top quark sector of LHT model, two singlet fields $T_{L_1}$ and $T_{L_2}$ (and their right-handed counterparts) are introduced to cancel the large radiative correction to the Higgs mass caused by the top quark loop. The Lagrangian containing the neutral Higgs boson interactions are given by:
\begin{eqnarray}
{\cal L}_t \supset - \lambda_1 f \left( \frac{s_\Sigma}{\sqrt{2}} \bar{t}_{L_+} t^{'}_R + \frac{1+c_\Sigma}{2}\bar{T}^{'}_{L_+}t^{'}_R \right) -\lambda_2 f (\bar{T}^{'}_{L_+}T^{'}_{R_+} + \bar{T}^{'}_{L_-}T^{'}_{R_-}) + ~{\rm h.c.}
\end{eqnarray}
where $c_\Sigma = \cos(\sqrt{2}h/f)$ and $s_\Sigma = \sin(\sqrt{2}h/f)$. We have defined the $T$-parity eigenstates as $t_{L_+}=(t_{L_1}-t_{R_1})/\sqrt{2}$, $T^{'}_{L_\pm}=(T_{L_1} \mp T_{L_2})/\sqrt{2}$ and $T^{'}_{R_\pm}=(T_{R_1} \mp T_{R_2})/\sqrt{2}$. Note that $T$-odd Dirac fermion $T_- \equiv (T^{'}_{L_-}, T^{'}_{R_-})$ does not have the tree level Higgs boson interaction, and thus it does not contribute to the Higgs mass at one-loop level.  The two $T$-even combinations $(t_{L_+},t^{'}_R)$ and $(T^{'}_{L_+},T^{'}_{R_+})$ are mixed as:
\begin{equation}
{\cal L}^{T-even}_{ mass} = - (\bar{t}_{L_+}~\bar{T}^{'}_{L_+})\, {\cal M} \,\left( \begin{tabular}{c} $t^{'}_R$ \\$T^{'}_{R_+}$ \end{tabular} \right) \,+\,~{\rm h.c.} \,,
\end{equation}
with
\begin{equation}\label{mass}
{\cal M} \,= \left( \begin{tabular}{cc} $\frac{\lambda_1 f}{\sqrt{2}} s_\Sigma$ & $0$ \\ $\frac{\lambda_1 f}{2} (1+c_\Sigma)$ & $\lambda_2 f$ \end{tabular} \right)\,.
\end{equation}
The mass matrix Eq.(\ref{mass}) can be diagonalized by defining the linear combinations,
\begin{eqnarray}
t_L &=& \cos\beta \,t_{L_+} - \sin\beta \,T^{'}_{L_+}, \quad T_{L_+} = \sin\beta \,t_{L_+} +\cos\beta \,T^{'}_{L_+}\nonumber \\
t_R &=& \cos\alpha \,t^{'}_R - \sin\alpha \,T^{'}_{R_+}, \quad T_{R_+} = \sin\alpha \,t^{'}_R + \cos\alpha \,T^{'}_{R_+}\ \label{combination}
\end{eqnarray}
where we used the dimensionless ratio $R=\lambda_1/\lambda_2$ to define the mixing angles $\alpha$ and $\beta$,
\begin{eqnarray}
\sin\alpha=\frac{R}{\sqrt{1+R^2}}, \quad \sin\beta=\frac{R^2}{1+R^2}\frac{v}{f}.
\end{eqnarray}
The $T$-even Dirac fermion $T_+ \equiv (T_{L_+}, T_{R_+})$, is responsible for canceling the quadratic divergence to the Higgs mass induced by the top quark loop.

Both $T_{+}$ and $T_{-}$ acquire a mass of order $f$ from a Yukawa-like Lagrangian. The masses of the top quark and its partners are given at $\mathcal O(v^{2}/f^{2})$ by
\begin{eqnarray}
&&m_t=\frac{\lambda_2 v R}{\sqrt{1+R^2}} \left[ 1 + \frac{v^2}{f^2} \left(
-\frac{1}{3} + \frac{1}{2} \frac{R^2}{(1+R^2)^2} \right)
\right]\nonumber \\
&&m_{T_{+}}=\frac{f}{v}\frac{m_{t}(1+R^2)}{R}\left[1+\frac{v^{2}}{f^{2}}\left(\frac{1}{3}-\frac{R^2}{(1+R^2)^2})\right)\right] \nonumber \\
&&m_{T_{-}}=\frac{f}{v}\frac{m_{t}\sqrt{1+R^2}}{R}\left[1+\frac{v^{2}}{f^{2}}\left(\frac{1}{3}-\frac{1}{2}\frac{R^2}{(1+R^2)^2})\right)\right]\label{Tmass}
\end{eqnarray}
where $R$ and $\lambda_2$ are considered to be free parameters. However, by using the measured top quark mass, we can fix $\lambda_2$ for a given $(f, R)$. Therefore, the only $f$ and $R$ are the free parameters in the top partner sector. On the other hand, $R \lesssim 3.3$ is required by the tree level unitarity limit of the $J=1$ partial-wave amplitudes in the coupled system of $(t\bar{t}, T_+ \bar{T}_+, b\bar{b}, W^+W^-, Zh)$ states \cite{cp-1}. From Eq.~\ref{Tmass}, we can see that the $T$-even top partner is always heavier than the $T$-odd heavy top partner, but it has more interesting phenomenological signatures due to the complicated decay modes. Besides, it is directly related with the electroweak naturalness of the LHT model. So, we will focus on the $T$-even top partner in this work.

Since there are usually two possible ways (they are denoted as Case A and Case B \cite{cp-2}) to construct the $T$-invariant Lagrangians of the Yukawa interactions of the down-type quarks and charged leptons, we will study both cases in our following parameter space scan. Up to $\mathcal{O}\left( v_{SM}^4/f^4 \right)$, the ratios of the Higgs couplings with the down-type quarks $g_{hd\bar{d}}$ with respect to the SM prediction $g^{SM}_{hd\bar{d}}$ can be expressed as,
\begin{eqnarray}
    \frac{g_{h \bar{d} d}}{g_{h \bar{d} d}^{SM}} &=& 1-
        \frac{1}{4} \frac{v_{SM}^{2}}{f^{2}} + \frac{7}{32}
        \frac{v_{SM}^{4}}{f^{4}} \qquad \text{Case A} \nonumber \\
    \frac{g_{h \bar{d} d}}{g_{h \bar{d} d}^{SM}} &=& 1-
        \frac{5}{4} \frac{v_{SM}^{2}}{f^{2}} - \frac{17}{32}
        \frac{v_{SM}^{4}}{f^{4}} \qquad \text{Case B}.
    \label{hdd}
\end{eqnarray}

\section{Constraints on the LHT model}
In our calculations, we assume a common Yukawa coupling $\kappa$ for all the mirror fermions and scan over the free parameters $\kappa$, $f$ and $R$ within the following region,
\begin{eqnarray}
500 ~{\rm GeV} \leqslant f \leqslant 2000 ~{\rm GeV}, \quad 0.1 \leqslant R \leqslant 3.3, \quad 0.6 \leqslant \kappa \leqslant 3.
\end{eqnarray}
where $\kappa \geqslant 0.6$ is from the constraint of the search for the monojet events at the LHC Run-1 \cite{lht-tonini}. Our scan approach is based on the frequentist theory, which has been used in our previous works \cite{lht-yang}. For a set of observables $\{ {\cal O}_i \}$, the experimental measurements are assumed to be Gaussian distributed with the mean value ${\cal O}^{exp}_i$ and error $\sigma_i$. The $\chi^2$ can be defined as $\chi^2 = \displaystyle{\sum_{i}^{N}}\frac{({\cal O}^{th}_i-{\cal O}^{exp}_i)^2}{{\sigma_i}^2}$. The likelihood $\cal{L}\equiv$exp$[-\displaystyle{\sum} \chi_{i}^{2}]$ for each point in the parameter space is calculated with the $\chi^2$ statistics as a sum of individual contributions from the latest experimental constraints. The confidence regions are evaluated by the profile-likelihood method from the values of $\delta\chi^2 \equiv -2 \ln({\cal L} / {\cal L}_{max})$. For three dimension scan, 1$\sigma$($2\sigma$) range is given by $\delta\chi^2 =3.53(8.02)$. We construct the likelihood function $\mathcal{L}$ by using the following constraints:

\begin{itemize}
\item[(1)] The electroweak precision observables: $S$, $T$ and $U$. In the LHT model, the top partner can correct the propagators of the electroweak gauge bosons at one-loop level. Meanwhile, due to the composite nature of the Higgs boson, the $S$ and $T$ parameters are modified by the deviation of the Higgs gauge couplings $hVV$ from the SM prediction \cite{lht-ewpos}. Besides, the UV operators can contribute to the $S$ and $T$ parameters \cite{zyhan}. The couplings of the UV operators are set as $c_{s} = c_{t} = 1$ \cite{lht-tonini}. We use the experimental values of $S$, $T$ and $U$ from Ref. \cite{pdg}.

\item[(2)] $R_{b}$. In the LHT model, the new mirror fermions and new gauge bosons can contribute to the $Zb\bar{b}$ coupling and give the corrections to the $R_{b}$ at one-loop level \cite{lht-rb-2}. The final combined result from the LEP and SLD measurements show $R_{b} = 0.21629\pm 0.00066$ \cite{pdg}, which is consistent with the SM prediction $R_{b}^{SM} = 0.21578^{+0.0005} _{-0.0008}$. In our work \cite{lht-rb-2}, it was also found that the LHT model can only slightly alleviate the tension between the $A_{FB}^b$ measurement and its SM prediction since the the correction of the new particles to $Zb\bar{b}$ couplings is mainly on the left-handed coupling.

\item[(3)] Higgs data. The signal strength of one specific analysis from a single Higgs boson can be given by
\begin{equation}
\mu = \sum_{i} c_i\omega_i,
\label{Eq:mu}
\end{equation}
where the sum runs over all channels used in the analysis. For each channel, it is characterized by one specific production and decay mode. The
individual channel signal strength can be calculated by
\begin{equation}
c_i=\frac{\left[\sigma\times BR\right]_i}{\left[\sigma_{SM}\times BR_{SM}\right]_i},
\label{Eq:ci}
\end{equation}
and the SM channel weight is
\begin{equation}
\omega_i=\frac{\epsilon_i\left[\sigma_{SM}\times BR_{SM}\right]_i}{\sum_j\epsilon_j\left[\sigma_{SM}\times BR_{SM}\right]_j}.
\label{Eq:omega}
\end{equation}
where $\epsilon_i$ is the relative experimental efficiencies for each channel. But these are rarely quoted in experimental publications. In this case, all channels considered in the analysis are treated equally, i.e. $\epsilon_i=1$. We confront the modified Higgs-gauge interactions  $hVV$, $hgg$ and $h\gamma\gamma$ within our model with the Higgs data by calculating the $\chi^{2}_{H}$ of the Higgs sector using the public package \textsf{HiggsSignals-1.4.0} \cite{higgssignals}, which includes the available Higgs data sets from the ATLAS, CMS, CDF and D0 collaborations. We choose the mass-centered $\chi^2$ method in the package \textsf{HiggsSignals}.
\end{itemize}
On the other hand, the current LHC direct searches for the multi-jet with the transverse missing energy can also produce the bounds on the parameter space of the LHT model. However, they are not strong enough to push the exclusion limits much beyond the indirect constraints \cite{lht-tonini}. So in our scan, we consider the above indirect constraints. It should be mentioned that since the SM flavor symmetry is broken by the extension of the top quark sector, the mixing between top partner and down-type quark can induce flavor changing neutral current processes at one-loop level \cite{lh-flavor}. Among them, the most sensitive one is the rare decay $B_s \to \mu^+\mu^-$. The latest combined result from the CMS and LHCb measurements has shown $Br_{exp}(B_s \to \mu^+\mu^-)=(2.9 \pm 0.7)\times 10^{-9}$ \cite{bs-exp}, which is well consistent with the SM prediction $Br_{SM}(B_s \to \mu^+ \mu^-)=(3.56\pm0.30) \times 10^{-9}$ \cite{bs-th}. We checked our samples and found that the constraints from $B_s \to \mu+\mu-$ can be easily satisfied within the current uncertainty due to the heavy mirror quark contributions being small.

\begin{figure}[ht]
\centering
\includegraphics[width=3.5in]{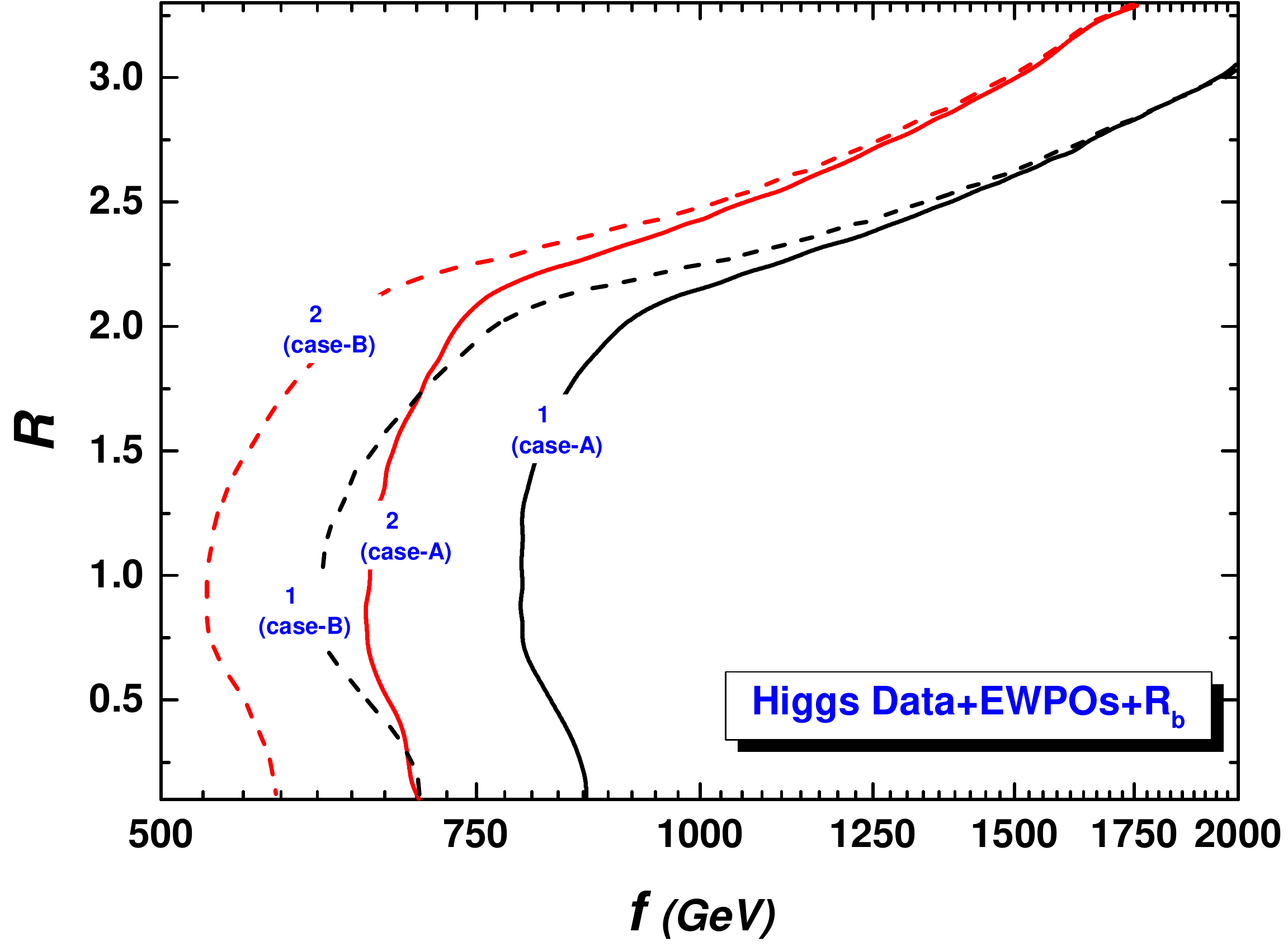}
\caption{Excluded regions (above each contour) in the $R$ versus $f$ plane of the LHT model for Case A and Case B, where the parameter $\kappa$ is marginalized over.}
\label{constraints}
\end{figure}
In Fig.~\ref{constraints}, we present the excluded regions (above each contour) by the global fit of the Higgs data, EWPOs and $R_b$ in the $R$ versus $f$ plane of the LHT model for Case A and Case B, where the parameter $\kappa$ is marginalized over. From the Fig.~\ref{constraints}, we can see that the symmetry breaking scale $f$ has been excluded up to about 589 (518) GeV at $2\sigma$ level for Case A (B) \footnote{These results are slightly weaker than Refs.~\cite{lht-tonini,lht-yang} because of the marginalization over the parameter $\kappa$.}, which corresponds to $m_{T_+}=829(730)$ GeV for $R=1$. The reason for lower bound on $f$ in Case B compared to Case A is that the reduced bottom Yukawa coupling in Case B is smaller than that in Case A (cf. Eq.~\ref{hdd}), which leads to a higher suppression of the branching ratio of $h \to b\bar{b}$, and hence an enhancement of all other decay rates. Such results are more consistent with the current LHC Higgs data, in particular with the ATLAS measurements, where a generic enhancement in the non-fermionic decays of the Higgs is reported.

\section{$T_{+}(\to t h)j$ production in the diphoton channel at the LHC}

\subsection{Single and Pair Production of $T_+$}
\begin{figure}[ht]
\centering
\includegraphics[width=4in]{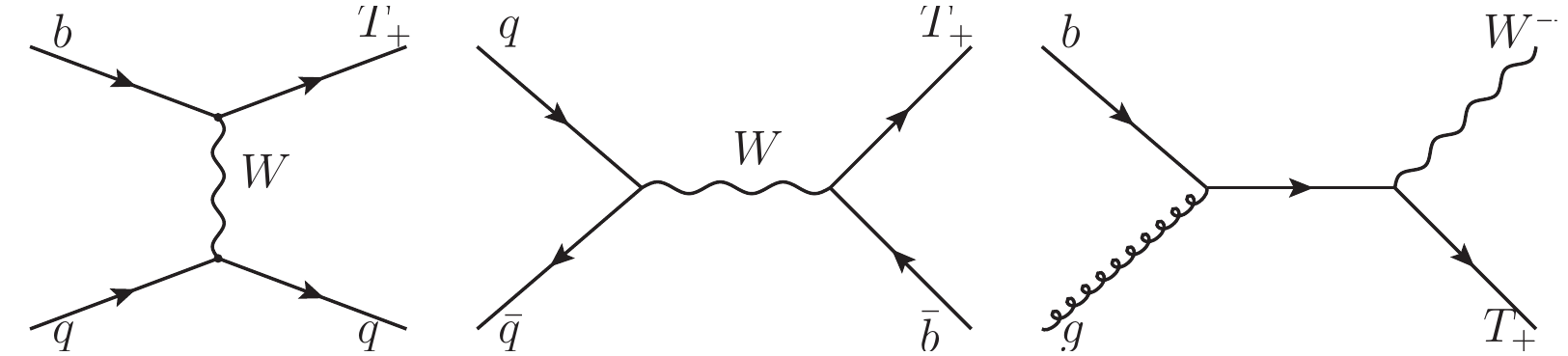}
\caption{Feynman diagrams for the single $T$-even top partner production at the LHC.}
\label{feyn}
\end{figure}
At the LHC, the single production of the $T$-even top partner is induced by the electroweak interaction and proceeds through the processes depicted in Fig.~\ref{feyn}. We investigate the $t$-channel process $qb \to T_+ q$ via $W$-exchange, which has the largest cross section among the three single production modes. In our numerical calculations, we use the input parameters of the SM as \cite{pdg}
\begin{eqnarray}
m_t&=&173.07{\rm ~GeV}, ~~m_W = 80.385~, ~~m_{Z}=91.19 {\rm~GeV},\nonumber
\\ && \sin^{2}\theta_W=0.2228, ~~\alpha(m_Z)^{-1}=127.918.
\end{eqnarray}
Besides, the Higgs mass is taken as $m_h=125.09$ GeV \cite{higgs-mass} and the CKM matrix is assumed to be diagonal. We use CTEQ6L as the parton distribution functions (PDF) in the calculation of the hadronic level cross section of the process $qb \to T_+ j$ \cite{cteq}. The renormalization scale $\mu_R$ and factorization scale $\mu_F$ are chosen to be $\mu_R=\mu_F=m_{T_+}/2$. Since the mirror fermion Yukawa coupling $\kappa$ is independent of the single top partner production, we fix $\kappa=2$ for simplicity.

\begin{figure}[ht]
\centering
\includegraphics[width=3.5in]{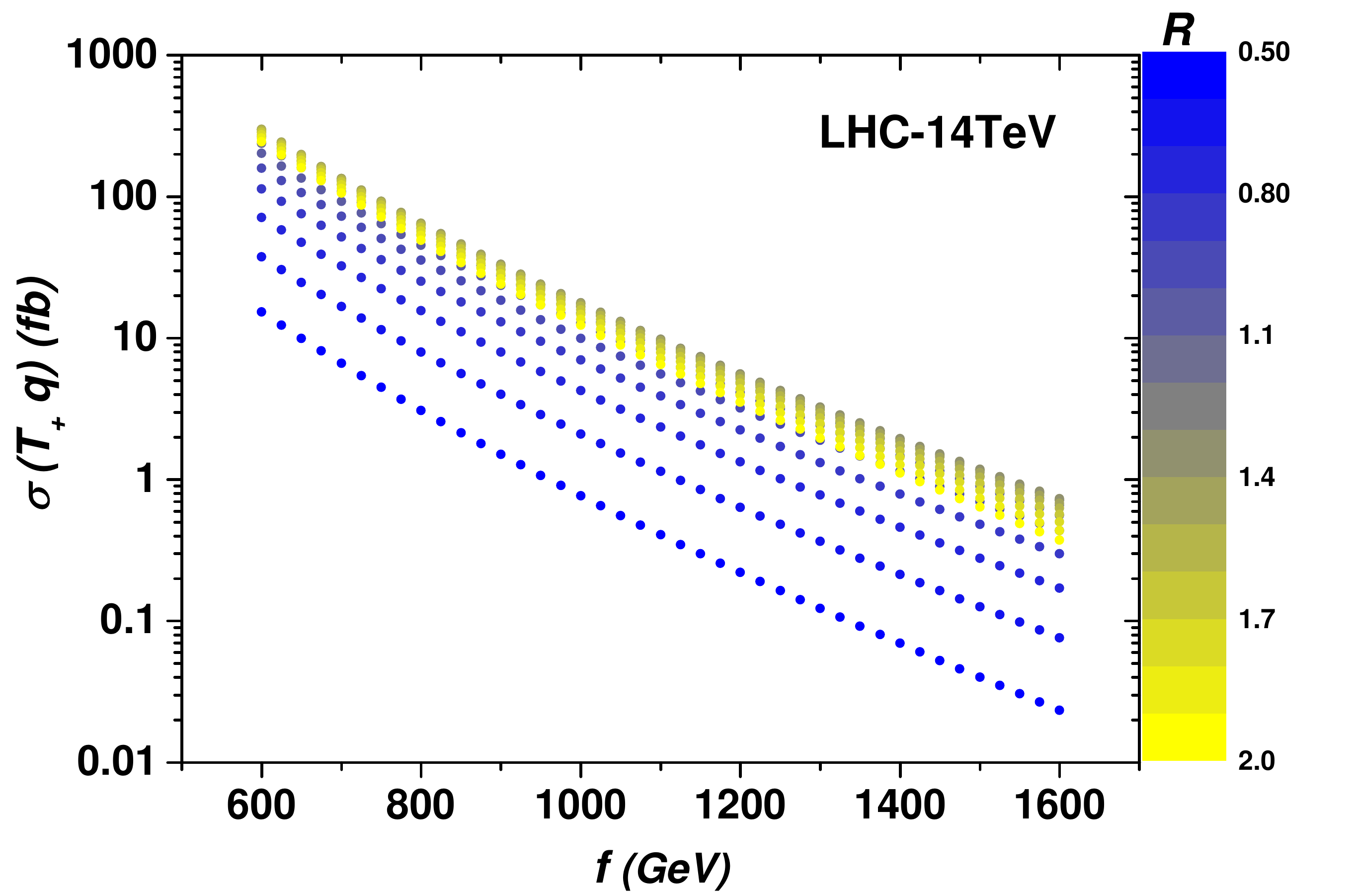}
\caption{The dependence of the cross section of the process $qb \to T_+ q$ on the symmetry breaking scale $f$ and the mixing parameter $R$ in the LHT model at 14 TeV LHC.}
\label{cx}
\end{figure}
In Fig.~\ref{cx}, we show the dependence of the cross section of the process $qb \to T_+ q$ on the symmetry breaking scale $f$ and the mixing parameter $R$ in the LHT model at 14 TeV LHC, where the contribution of the charge-conjugate process $\bar{T}_+ q$ is not included. Since the coupling of $T_+bW$ is proportional to the ratio $R^2/(1+R^2)$, the cross section of the process $qb \to T_+ q$ will become larger with the increase of $R$. However, the mass of $T_+$ also depend on the mixing parameter $R$ (cf. Eq.~\ref{Tmass}). Therefore, from Fig.~\ref{cx}, we can see that the cross section of the process $qb \to T_+ q'$ maximally reach about 295 fb when $f=600$ GeV and $R=1.5$. With the increase of the $T_+$ mass, its cross section decreases rapidly and is less than 1 fb if $f \geq 920~(1370)$ GeV for $R=0.5~(2.0)$.

\subsection{LHC observability of $T_+(\to th)j \to t(\to b \ell^+ v_\ell)h(\to \gamma\gamma)j$}
In the LHT model, decay channels of the $T$-even top partner include $T_+ \to bW$, $T_+ \to tZ$, $T_+ \to th$ and $T_+ \to T_- A_H$. Due to the Goldstone-boson equivalence theorem, we can have the branching ratios relationship $\frac{1}{2}Br(bW) \simeq Br(tZ) \simeq Br(th)$ in the limit $f \gg v$. In the following calculations, we perform the Monte Carlo simulation and explore the sensitivity of $T_+j$ production at 14 TeV LHC through the channel,
\begin{equation}\label{signal-process}
pp \to T_+(\to th)j \to t(\to b \ell^+ v_\ell)h(\to \gamma\gamma)j.
\end{equation}
which features that two photons in the final states appear as a narrow resonance centered around the Higgs boson mass. Therefore, the SM backgrounds to the Eq.~(\ref{signal-process}) include the resonant and the non-resonant processes. For the former, they have a a Higgs boson decaying to diphoton in the final states, such as $t\bar{t}h$ and $thj$ productions. For the latter, they are the diphoton events produced in association with top quarks, such as $tj\gamma\gamma$ and $t\bar{t}\gamma\gamma$ productions.

We merge the effective interaction $h\gamma\gamma$ into the LHT model file \cite{lht-tonini} that are generated by the package \textsf{FeynRules} \cite{feynrules}. We calculate the partial decay widths of $T_+$ with the \textsf{Madgraph5} \cite{mad5} and feed their values into the parameter card. The branching ratio of $h \to \gamma\gamma$ is normalized to its LHT model prediction. We generate the signal and background events by using \textsf{MadGraph5} and perform the parton shower and the fast detector simulations with \textsf{PYTHIA} \cite{pythia} and \textsf{Delphes} \cite{delphes}, respectively. When generating the parton level events, we assume $\mu_R=\mu_F$ to be the default event-by-event value. We cluster the jets by choosing the anti-$k_t$ algorithm with a cone radius $\Delta R=0.5$ \cite{anti-kt}. The $b$-jet tagging efficiency is parameterized as a function of the transverse momentum and rapidity of the jets \cite{cms-b}. In the simulation, we generate 1.2 million events for the signal and each background, respectively. The cross section of $t\bar{t}h$ and $thj$ production are normalized to their NLO values \cite{tth,thj}.

\begin{figure}[ht]
\centering
\includegraphics[width=2.5in,height=2in]{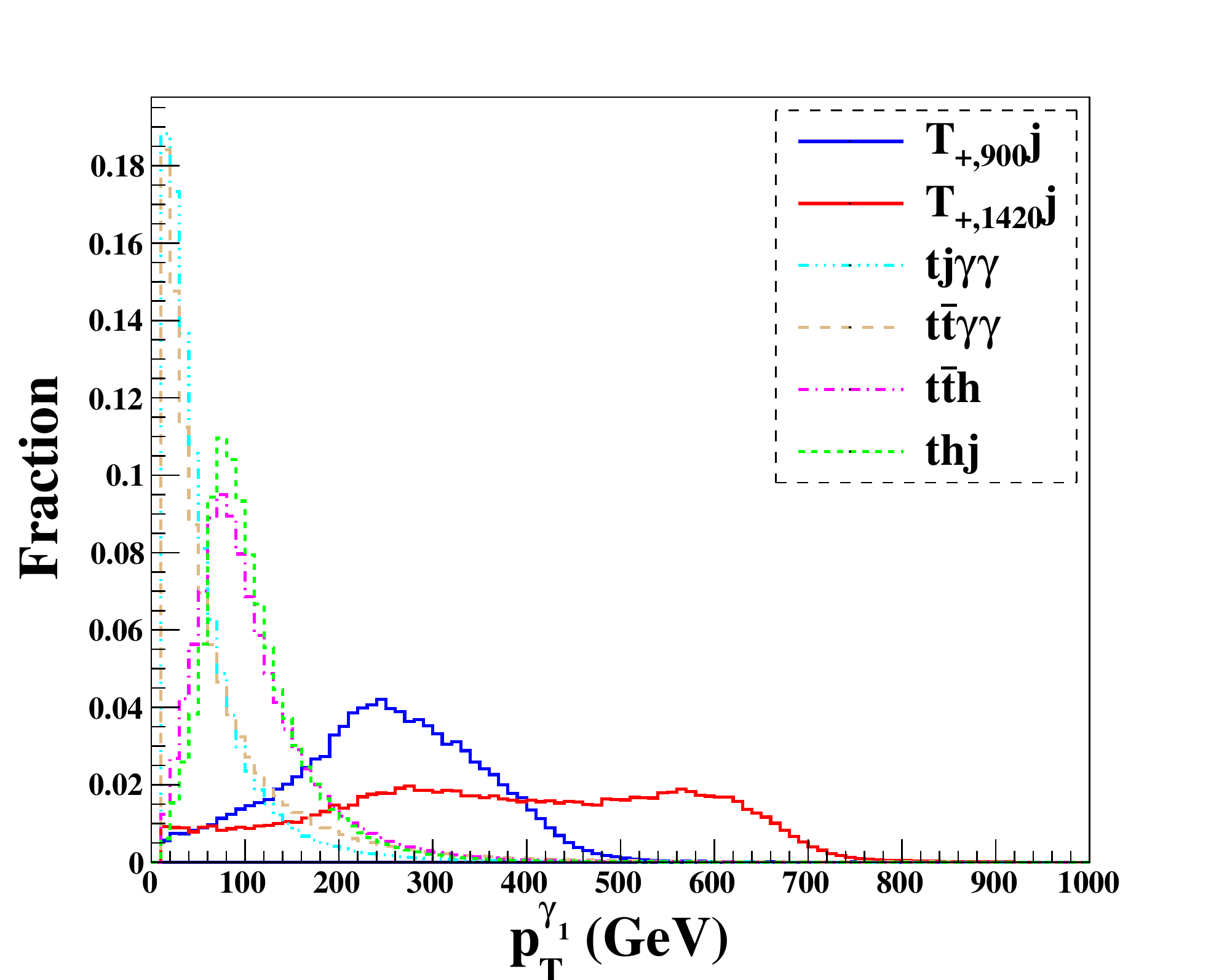}
\includegraphics[width=2.5in,height=2in]{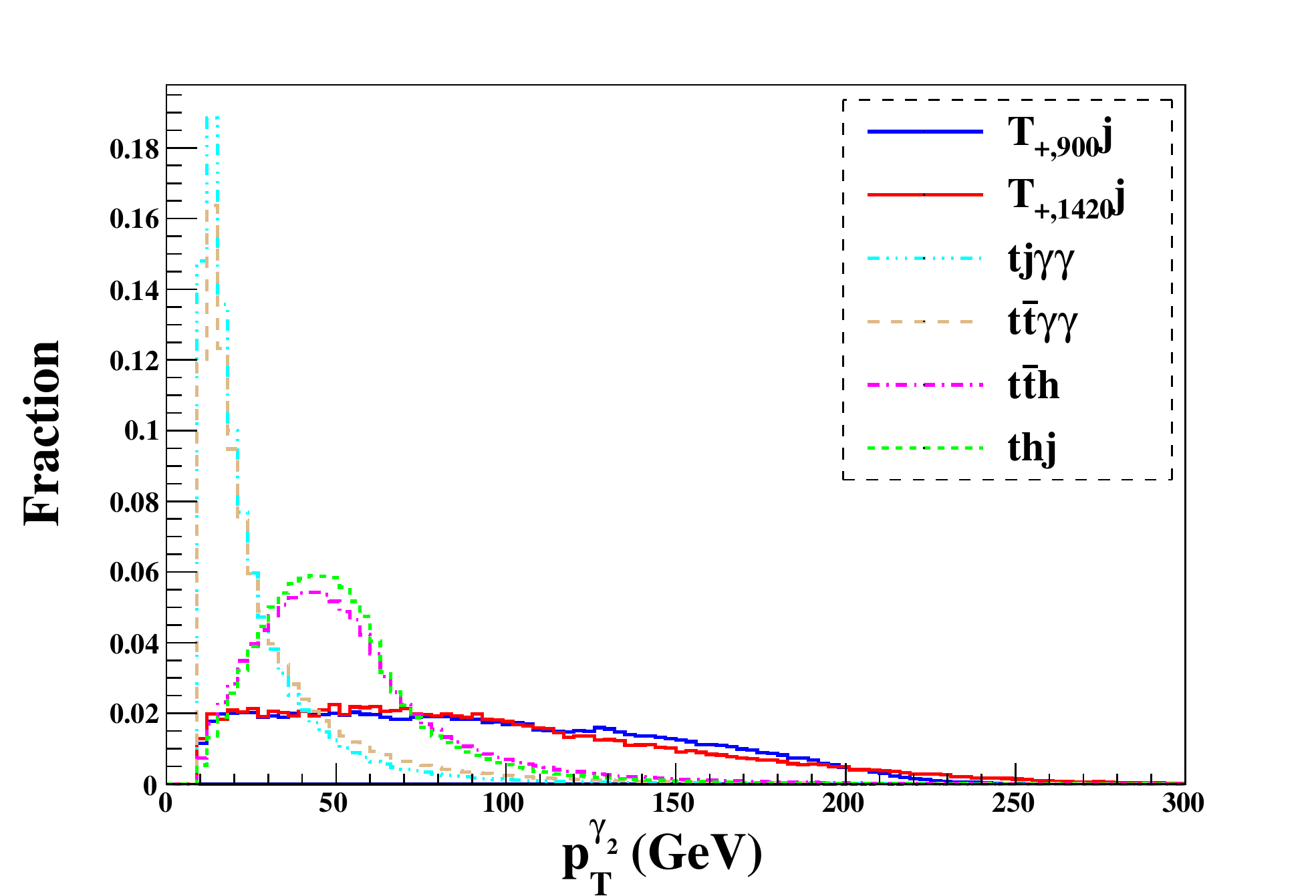}
\includegraphics[width=2.5in,height=2in]{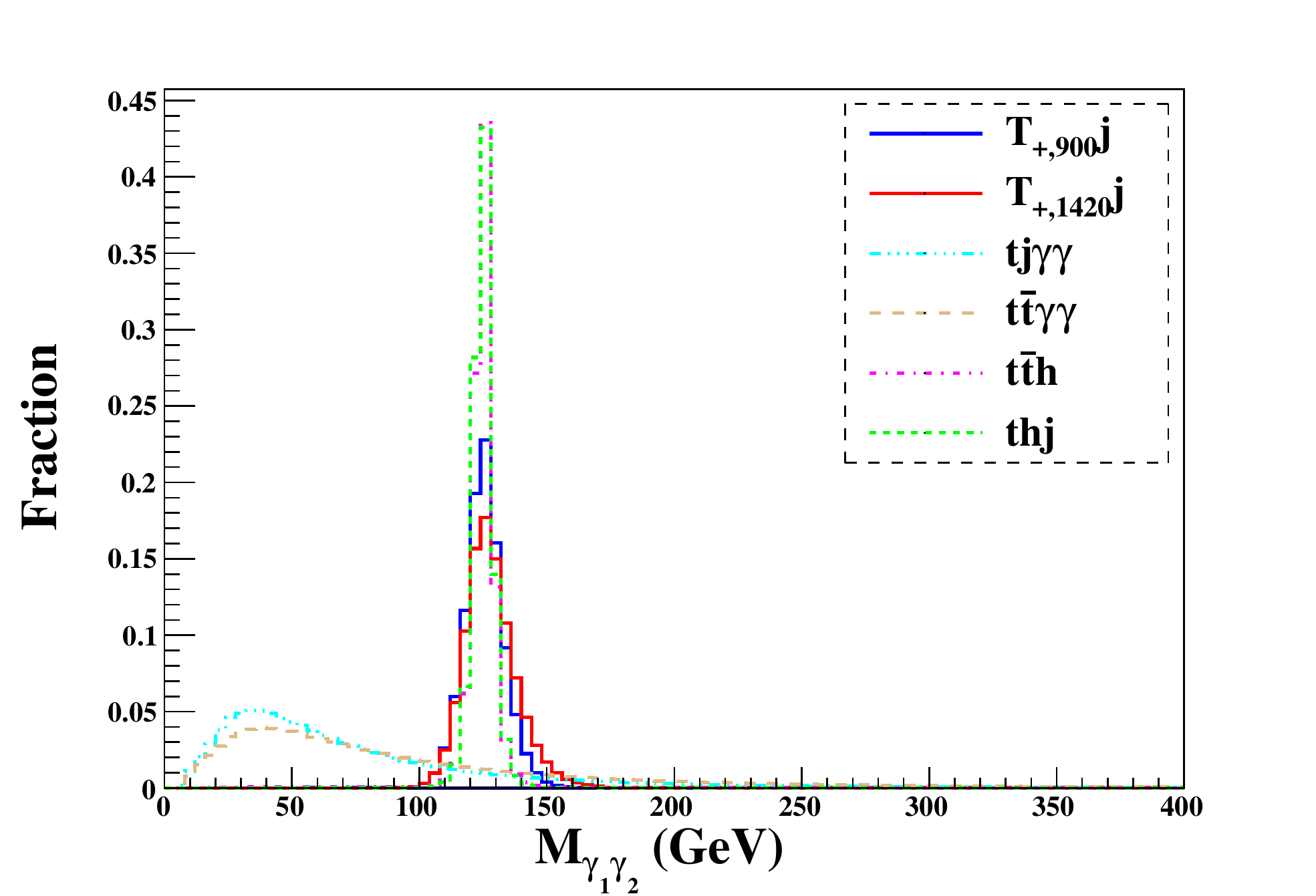}
\includegraphics[width=2.5in,height=2in]{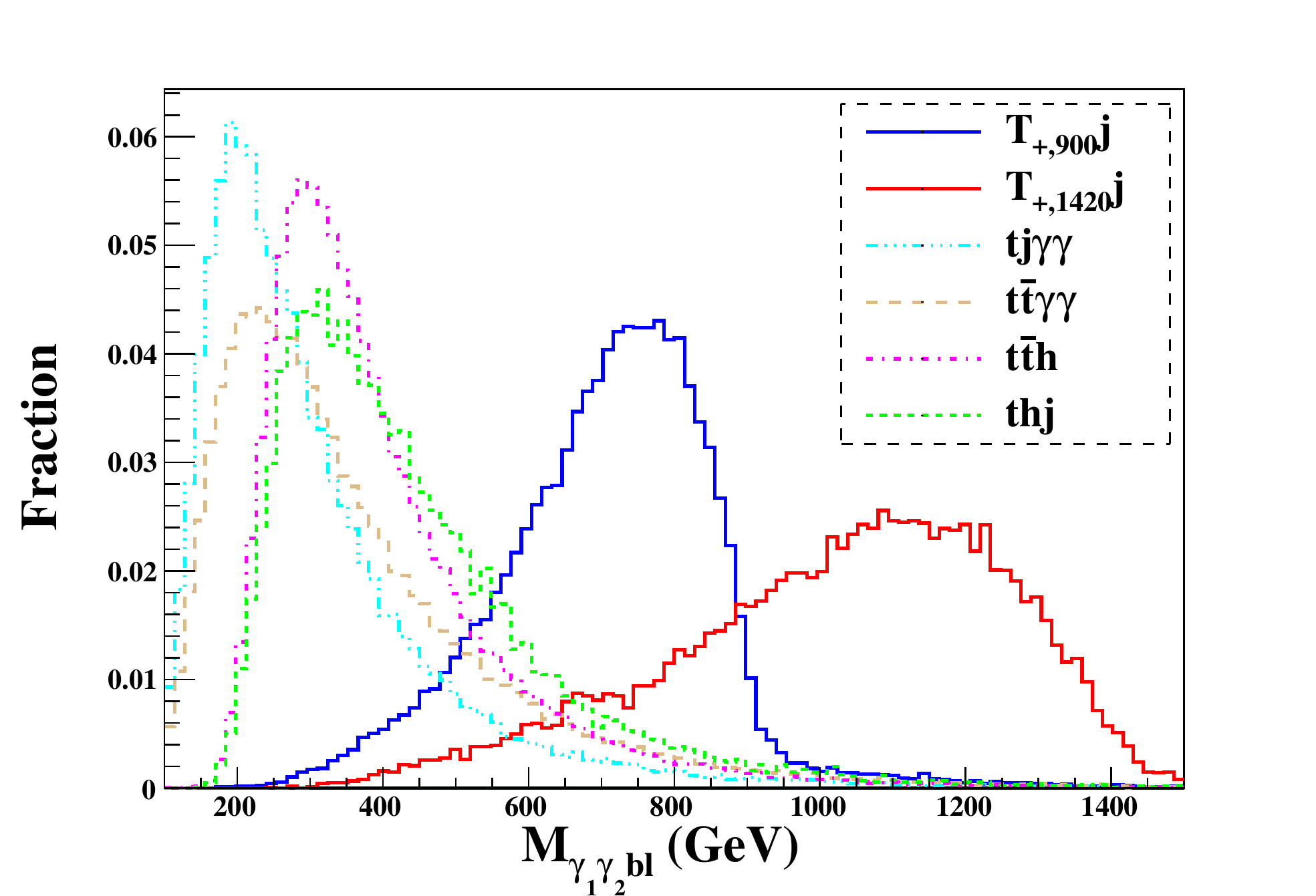}
\caption{The event fractions of the signals and the backgrounds with respective to the transverse momentum distributions $p^{\gamma^1}_T$ and $p^{\gamma^2}_T$ (upper panel), the invariant mass distributions $m_{\gamma_1\gamma_2}$ and $m_{\gamma_1\gamma_2 b \ell}$ (lower pannel) at 14 TeV LHC. The two signal benchmark points correspond to $(f,R)=(500,0.5)$ and $(800,0.5)$, which give the $T$-even top partner mass $m_{T_+}=900$ GeV and 1420 GeV, respectively.}
\label{dis}
\end{figure}

In Fig.~\ref{dis}, we show the transverse momentum distributions $p^{\gamma^1}_T$ and $p^{\gamma^2}_T$ (upper panel), the invariant mass distributions $m_{\gamma_1\gamma_2}$ and $m_{\gamma_1\gamma_2 b \ell}$ of the signals and backgrounds at 14 TeV LHC. The two signal benchmark points correspond to $(f,R)=(500,0.5)$ and $(800,0.5)$, which give the $T$-even top partner mass $m_{T_+}=900$ GeV and 1420 GeV, respectively. Since the Higgs boson is boosted in the heavy top partner decay, the two photons from the Higgs decay in the signals have the harder $p_T$ spectrum than the backgrounds. In comparison with the hadronic decay of the Higgs boson, the $\gamma\gamma$ channel has the good resolution on the $\gamma\gamma$ resonance. From Fig.~\ref{dis}, we can see that the spreading of the $\gamma\gamma$ invariant mass peak at $m_h$ for the signals and the resonant backgrounds are relatively small. A narrow invariant mass window, such as $m_{\gamma\gamma}-m_h <5$ GeV, will greatly reduce the non-resonant backgrounds $tj\gamma\gamma$ and $t\bar{t}\gamma\gamma$. Besides, the invariant mass distribution $m_{\gamma\gamma b\ell}$ has an endpoint round the mass of $m_{T_+}$, which can be used to remove the backgrounds effectively.

\begin{table*}[t!]
\begin{center}
\caption{The keeping efficiency $\epsilon_{cut-i}(i=1,2,3,4,all)$ of the background events $tj\gamma\gamma$, $tt\gamma\gamma$, $t\bar{t}h$, $thj$ and the signal event $T_+j$ with $f=500$ GeV and $R=1.5$ (corresponding to $m_{T_+}=770$ GeV) at 14 TeV LHC. \label{cutflow}}
\begin{ruledtabular}
\begin{tabular}{c||c|c|c|c||c}
&$tj\gamma\gamma$ &$tt\gamma\gamma$  &$t\bar{t}h$ &$thj$  &$T_{+,770}j$ \\\hline
$\sigma(pb)$ & 0.012 & 0.011 & 0.586 & 0.088 & 0.299\\\hline
$\epsilon_{cut-1}$ & 0.99 & 0.67 & 0.65 & 0.99 & 0.98 \\\hline
$\epsilon_{cut-2}$ & 0.21 & 0.20 & 0.28 & 0.25 & 0.51 \\\hline
$\epsilon_{cut-3}$ & 0.009 & 0.015 &  0.47 & 0.48 & 0.46 \\\hline
$\epsilon_{cut-4}$ & 0.25 & 0.31 & 0.24 & 0.13 & 0.29 \\\hline
$\epsilon_{cut-all}$   & $7.50\times 10^{-6}$   & $4.17\times 10^{-6}$   & $2.16 \times 10^{-4}$  & $8.04\times 10^{-4}$    & $1.4\times 10^{-2}$   \\
\end{tabular}
\end{ruledtabular}
\end{center}
\normalsize
\end{table*}

According to the above analysis, we require the events to satisfy the following criteria:
\begin{itemize}
  \item cut-1: there is exact one isolated lepton with $p_T(\ell_1) > 20$ GeV and exact one $b$-jet with $p_T(b_1) > 25$ GeV. At most two hard jets have $p_T(j_{1,2}) > 25$ GeV;
  \item cut-2: there are two photons with $p^{\gamma_1}_{T}>120$ GeV and $p^{\gamma_2}_{T}>70$ GeV;
  \item cut-3: the invariant mass of two photons $m_{\gamma_1\gamma_2}$ should be in the range of $m_h\pm5$ GeV;
  \item cut-4: the invariant mass of two photons, $b$-jet and lepton $m_{\gamma_1\gamma_2 b \ell}$ is greater than $m_{T_+}/2$.
\end{itemize}

In Table \ref{cutflow}, we present the keeping efficiency $\epsilon_{cut-i}(i=1,2,3,4,all)$ of the background events $tj\gamma\gamma$, $tt\gamma\gamma$, $t\bar{t}h$, $thj$ and the signal event $T_+j$ with $f=500$ GeV and $R=1.5$ under the above cut flow at 14 TeV LHC. From Table \ref{cutflow}, we can see that the jet multiplicity selection $N_j \leq 2$ (i.e. cut-1) suppresses the the backgrounds involving $t\bar{t}$ final states, such as $t\bar{t} \gamma\gamma$ and $t\bar{t}h$. All the backgrounds are greatly removed by the requirement of the two high $p_T$ photon (i.e. cut-2) since the photons in the signal are from the boosted Higgs boson in the heavy top partner decay. The non-resonant backgrounds $tj\gamma\gamma$ and $t\bar{t}\gamma\gamma$ are efficiently reduced by ${\cal O}(10^{-2})$ due to the Higgs mass cut (i.e. cut-3). The requirement of the high invariant mass $m_{\gamma_1\gamma_2 b \ell} > m_{T_+}/2$ (i.e. cut-4) can hurt the background $thj$ more than the signal. So after all cuts, the largest background is $t\bar{t}h$ because of its large cross section.

\begin{figure}[ht]
\centering
\includegraphics[width=3.5in]{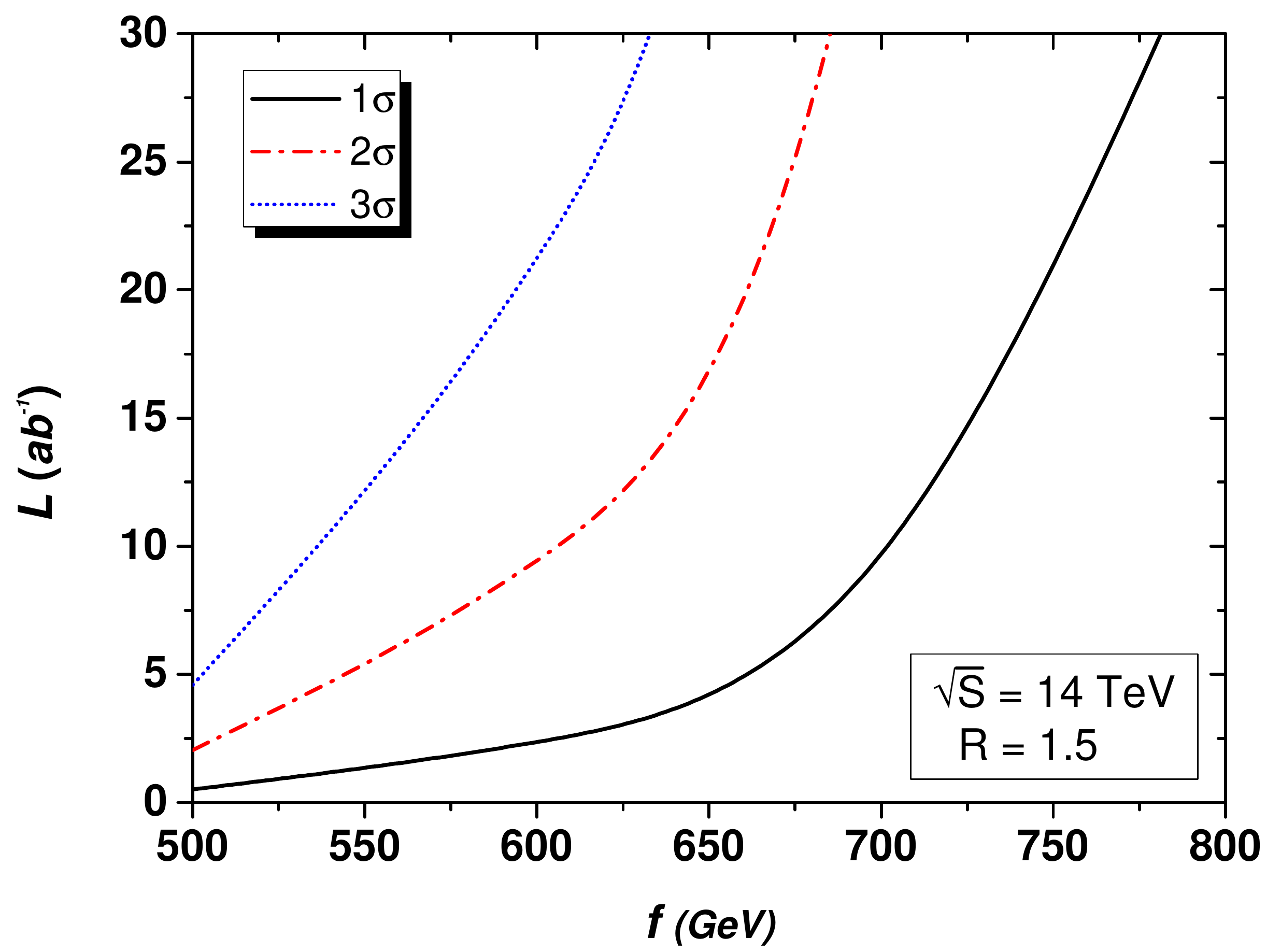}
\caption{Contour plots of the statistical significance $S/\sqrt{B}$ of $pp \to T_+j$ in the Higgs to diphoton channel at 14 TeV LHC on the plane of the integrated luminosity ${\cal L}$ versus the symmetry breaking scale $f$, where the mixing parameter $R$ is fixed at 1.5. The contribution of the charge conjugate process $pp \to \bar{T}_+ j$ has been included.}
\label{ss}
\end{figure}

In Fig.~\ref{ss}, we plot the contours of the statistical significance $S/\sqrt{B}$ of $pp \to T_+j$ in the Higgs to diphoton channel at 14 TeV LHC on the plane of the interated luminosity ${\cal L}$ versus the symmetry breaking scale $f$, where the mixing parameter $R$ is fixed at 1.5. The contribution of the charge conjugate process $pp \to \bar{T}_+ j$ has been included in the calculations. From Fig.~\ref{ss}, we see that the scale $f$ can be excluded up to 520 GeV ($m_{T_+}=800$ GeV) at $2\sigma$ level at 14 TeV LHC with the integrated luminosity ${\cal L}=3$ ab$^{-1}$. This is mildly better than the current indirect bound $m_{T_+}>730$ GeV. If the luminosity can reach about 10 ab$^{-1}$, the $2\sigma$ exclusion limit of the scale $f$ will be pushed up to 610 GeV ($m_{T_+}=936$ GeV). On the other hand, it is worth mentioning that the top quark from the decay of $T_+ \to th$ usually has the unbalanced polarization states because of the $T_+th$ coupling being,
\begin{equation}
C_{htT_+}=-im_t(\frac{1}{1+R^2}\frac{1}{f}P_R-R\frac{1}{v}P_L).
\end{equation}
Such a feature may lead to the different angular distributions of the top quark decay products from those of the dominant backgrounds $pp \to t\bar{t}h$ via QCD interaction and $pp \to thj$ via the SM electroweak interaction, which may be utilized to further distinguish the signal from its backgrounds \cite{pol-1,pol-2}.

\section{Conclusion}\label{section4}
In this paper, we firstly examined the parameter space of the LHT by considering the constraints from the Higgs data, the electroweak precision observables and $R_b$, and found that the mass of $T$-even top partner ($T_+$) should be heavier than 730 GeV. Then, under these constraints, we calculated the cross sections of the single $T_+$ production and the pair $T_+$ production and the various branching ratios of $T_+$. Finally, we
investigated the observability of the single $T_+$ production through the process $pp \to T_+ j$ with the sequent decay $T_+ \to th$ in the di-photon channel in the LHT model at the LHC. We found that the mass of $T_+$ can be excluded up to 800 GeV at $2\sigma$ level at 14 TeV LHC with the integrated luminosity ${\cal L}=3$ ab$^{-1}$, which is mildly better than the current bound from the indirect searches.

\acknowledgments
This work is partly supported by the Australian Research Council, by the National Natural Science Foundation of China (NNSFC) under grants Nos. 11275057, 11305049, 11375001 and 11405047, by Specialised Research Fund for the Doctoral Program of Higher Education under Grant No. 20134104120002 and by the Startup Foundation for Doctors of Henan Normal University under contract No.11112, by the China Postdoctoral Science Foundation under Grant No. 2014M561987 and the Joint Funds of the National Natural Science Foundation of China (U1404113).

\end{document}